%% file: dftau.tex
\documentclass[
 aps, pra,
 amsmath,amssymb,
 11pt,
 final,
tightenlines,
 twoside,
 twocolumn,
 nofloats,
nofootinbib,
 superscriptaddress,
showkeys,
showkeywords,
 ]
{revtex4-2}

\usepackage[T2A]{fontenc}
\usepackage[utf8]{inputenc}
\usepackage[english]{babel}
\usepackage{graphicx}% Include figure files
\usepackage{dcolumn}% Align table columns on decimal point
\usepackage{bm}% bold math

\usepackage{xcolor}
\usepackage{soul}
\usepackage{lineno}
\usepackage{changes}

\definecolor{green2}{rgb}{0.0,0.5,0}

\input{maik.rty}

\setcitestyle{authoryear,round}
\input{sao_cmd_author.tex}

\def\kms{km\,s$^{-1}${}}
\def\micron{$\mu$m{}}

\begin{document}

\selectlanguage{english}

%\ydk{}
%\titlerunning{}
%\authorrunning{}
%\toctitle{}
%\tocauthor{}

\keywords{\it stars: variables: T~Tauri, Herbig Ae/Be---stars: individual: DF~Tau---ISM: jets and outflows}

{\it Accepted by Astrophysical Bulletin}
\title{On the influence of component orbital motion \\ on the photometric variability of DF Tau}

\author{\firstname{M.~A.}~\surname{Burlak}}
\affiliation{Sternberg Astronomical Institute, M.V. Lomonosov Moscow State University, Moscow, 119234 Russia\\}
\author{\firstname{K.~N.}~\surname{Grankin}}
\affiliation{Crimean Astrophysical Observatory RAS, 298409 Nauchny, Republic of Crimea\\}
\author{\firstname{A.~V.}~\surname{Dodin}}
\author{\firstname{N.~V.}~\surname{Emelyanov}}
\author{\firstname{N.~P.}~\surname{Ikonnikova}}
\author{\firstname{Ya.~A.}~\surname{Lazovik}}
\author{\firstname{S.~A.}~\surname{Lamzin}}
\email{lamzin@sai.msu.ru}
\author{\firstname{B.~S.}~\surname{Safonov}}
\author{\firstname{I.~A.}~\surname{Strakhov}}
\affiliation{Sternberg Astronomical Institute, M.V. Lomonosov Moscow State University, Moscow, 119234 Russia\\}

        %%%%%%%%%%%%%%%%%%%%%%%%%%%%%%%%%%%%%%%%%%%%%%%%%%%%%%%%

\begin{abstract}

Based on the analysis of the long-term light curve of the young binary DF~Tau spanning approximately 125~years, we infer that its brightness variations are associated with changes in the accretion rate from the circumstellar protoplanetary disk onto the primary. We have also substantially improved the orbital parameters of DF~Tau, which enables us to align its secular light curve with the evolution of the binary’s component separation. The relationship between the long-term brightness variations and the orbital motion of DF~Tau, if present, appears to be inconsistent with theoretical predictions. Notably, similar discrepancies between theory and observations are also seen in other young binary systems.

Furthermore, the source of the polarized radiation in the optical range is found to be located at a distance of $\lesssim 0.5^{\prime\prime}$ from the star, with the polarization variability showing no dependence on the orbital phase.
\end{abstract}
                            %%%%%%%%%%%%%%%%%%%%%%%%%%%%%%%%%%%%%%%%%%%%
\maketitle

\section{Introduction}
 \label{sect:introduct}
 
Classical T Tauri stars (CTTS) are young ($t \lesssim 10^7$~yr), low-mass ($M \lesssim 2$~M$_\odot$) pre-main-sequence stars. Their activity is predominantly driven by magnetospheric disk accretion \citep{BBB-1988, Hartmann-2016}, which typically produces outflows in the form of a weakly collimated disk wind and, in some cases, collimated jets \citep{Bally-2016}.

The optical variability of CTTS on timescales ranging from tens of minutes to several days can arise from a variety of mechanisms: eclipses by gas-dust clouds, cool and/or hot spots on the stellar surface, non-stationary accretion, and/or chromospheric flares (see \citet{Herbst-94, Lamzin-2022, Lin-2023} and references therein).

At the same time, photographic \citep{Kholopov_eng-1970} and photoelectric \citep{Grankin-07} observations reveal that a number of CTTS exhibit smooth, wave-like variations in their mean brightness on timescales $\gtrsim 10$~yr. The long-period photometric variability of CTTS is likely driven by multiple mechanisms. For instance, \citet{Burlak-2025} proposed that the long-term brightness changes in the single star BP~Tau are linked to quasi-periodic modulations of the stellar magnetic field, while the systematic decline in brightness of T~Tau~N over the past decade is attributed to eclipses by the gas-dust disk surrounding the triple system T~Tau~N+Sa+Sb \citep{Beck-2025}.

In binary CTTS systems with orbital periods $P$ of several tens of years, long-period brightness variations may be attributed to modulation of the accretion rate due to tidal interactions between the companion and the primary star's disk and/or a circumbinary disk (CB-disk). Such modulation appears to be observed in the close binary system DQ~Tau \citep{Tofflemire-2017, Tofflemire-2025}, which has an orbital period of $P \approx 15\fd8$ and eccentricity $e = 0.57$, but is absent in ZZ~Tau ($P = 46.8 \pm 0.8$~yr, $e = 0.58 \pm 0.02$; \citealt{Belinski-2022}). In this context, the young binary DF~Tau is of particular interest: its orbital parameters are similar to those of ZZ~Tau, yet it exhibits long-period variability with a timescale comparable to its orbital period \citep{Lamzin-2001}.

DF~Tau (MHA~259-11) first attracted attention after \citet{Joy-1949} detected a bright H$\alpha$ emission line in its spectrum and proposed that it was a T~Tauri star. This hypothesis was further supported by the discovery of strong photometric variability ($\Delta m_{\text{pg}} > 2^{\text{m}}$) in DF~Tau \citep{Kholopov-Kurochkin-1951}. Due to the large astrometric uncertainty in the Gaia solution (RUWE $\approx 22$), we adopt the mean distance to the D4-North subgroup -- of which DF~Tau is a member -- as the distance to DF~Tau: $d = 140$~pc \citep{Krolikowski-2021, Kutra-2025}.
 
\citet{Chen-1990} discovered that DF~Tau is a binary system, which consists of two approximately equal-mass CTTSs \citep{Hartigan-Kenyon-2003}. According to \citet{Kutra-2025}, the effective temperatures of components A and B are $T_\text{eff} = 3640 \pm 100$~K and $3430 \pm 80$~K, respectively, and both stars possess global magnetic fields with strengths of $\approx 2.5$~kG. ALMA interferometric observations reveal that each component is surrounded by a gas-dust (protoplanetary) disk, but no CB disk has been detected \citep{Kutra-2025}.

It has long been known that intense outflows originate from the vicinity of DF~Tau \citep{Edwards-1994, HEG-1995, Lamzin-HST-2001, Nisini-2024}. These outflows manifest both as a bipolar jet (and associated Herbig--Haro objects, including HH~1266) and as a weakly collimated wind, which has formed a ring-shaped emission nebula with a radius of $\sim 2\times10^4$~au surrounding the binary system \citep{Dodin-2025}.

The orbit of DF~Tau was first determined by \citet{Thiebaut-1995}, who, using the earliest astrometric measurements, found an orbital period of $P = 84 \pm 12$~yr, a semi-major axis of $a = 190 \pm 30$~mas, and an eccentricity of $e = 0.8 \pm 0.3$. Based on these orbital parameters, \citet{Lamzin-2001} compared the long-term light curve of DF~Tau with the changing separation between the components and concluded that the rate of accretion onto DF~Tau~A is modulated in some manner by the orbital motion of its companion, DF~Tau~B. However, they noted that more precise orbital constraints are required for definitive conclusions. Since that publication, the orbit of DF~Tau has been significantly refined, and additional photometric data have been accumulated. We therefore revisit the question of whether the century-scale photometric variability of the system can be explained by accretion rate modulation driven by orbital motion.

This paper is organized as follows. In Section~2, we describe the observational data underlying our study; in Section~3, we present the results; and in Section~4, we provide their interpretation. The main conclusions are summarized in the Conclusion.

                %%%%%%%%%%%%%%%%%%%%%%%%%%%%%%%%%%%%%%%%%%%%%%%
                 
\section{Observational data}
 \label{sect:observation}
 
Between 2022 and 2024, we carried out photometric observations of DF~Tau using the 60-cm telescope at the Caucasus Mountain Observatory (CMO) of SAI MSU \citep{Berdnikov2020}, equipped with a CCD camera and a set of $UBVRI$ filters following the Bessel--Cousins system \citep{Bessel-1990}.

Polarimetric observations of DF~Tau were conducted on 2~December~2023 at 22:31~UT using a speckle polarimeter mounted on the 2.5-m telescope at CMO. Observations were performed in the $I_\mathrm{c}$ band using a two-beam polarimetry mode with a continuously rotating half-wave plate. The instrument was placed at the Nasmyth 2 focus, yielding a field of view of $5^{\prime\prime}\times5^{\prime\prime}$. The data reduction procedure, including correction for instrumental polarization, is described in \citet{Safonov-2017}.

For the orbital solution of DF~Tau, we used the same observational data as \citet{Kutra-2025} --- see Table~\ref{tab:orbit-dat} in the Appendix.
%Table~\ref{tab:orbit-dat}, the full version of which is provided in the electronic supplement to this paper.

               %%%%%%%%%%%%%%%%%%%%%%%%%%%%%%%%%
                
\section{Results}

\subsection{Orbit of DF~Tau}  
\label{sec:orbit}

As noted in the Introduction, the orbit of DF~Tau was first determined by \citet{Thiebaut-1995}. As new observational data became available, the orbital parameters were progressively refined (see \citealt{Schaefer-2006} and references therein). In particular, the orbital period decreased by a factor of $\sim$1.5 compared to the initial estimate. However, by the time of the studies by \citet{Schaefer_2014} and \citet{Allen-2017}, discrepancies among the derived parameters were no larger than the uncertainties of the respective methods -- see Table~\ref{tab:orbit-params}. In this context, we find it surprising that some orbital parameters, particularly the inclination of the orbit to the line of sight $i$, determined by \citet{Kutra-2025} using the most recent data, have changed substantially.

For this reason, we have recomputed the orbital parameters of DF~Tau using the same observational data as \citet{Kutra-2025}, employing two distinct methods. The first method (M1 in Table~\ref{tab:orbit-params}) is a least-squares fitting procedure \citep{Emelya-2020}, in which observational uncertainties were incorporated as weights in the formulation of the condition equations.

As the second method (M2 in Table~\ref{tab:orbit-params}), we employed the \verb|orbitize!| package for Python \citep{2020AJ....159...89B} in combination with the \verb|ptemcee| package \citep{2016MNRAS.455.1919V}, which implements a parallel-tempered affine-invariant Markov Chain Monte Carlo (MCMC) algorithm. The algorithm was run with the following parameters: 20 temperature levels\footnote{In MCMC terminology, ``temperature'' refers to a parameter that controls the exploration of parameter space: higher temperatures allow broader sampling across low-probability regions, while lower-temperature chains focus on refining samples near local likelihood maxima.}, $10^3$ walkers, $1.5\times10^7$ total steps, a burn-in of $10^4$ steps per chain, and a thinning factor of 100.

\begin{figure*}
   \centering
   \includegraphics[width=\hsize]{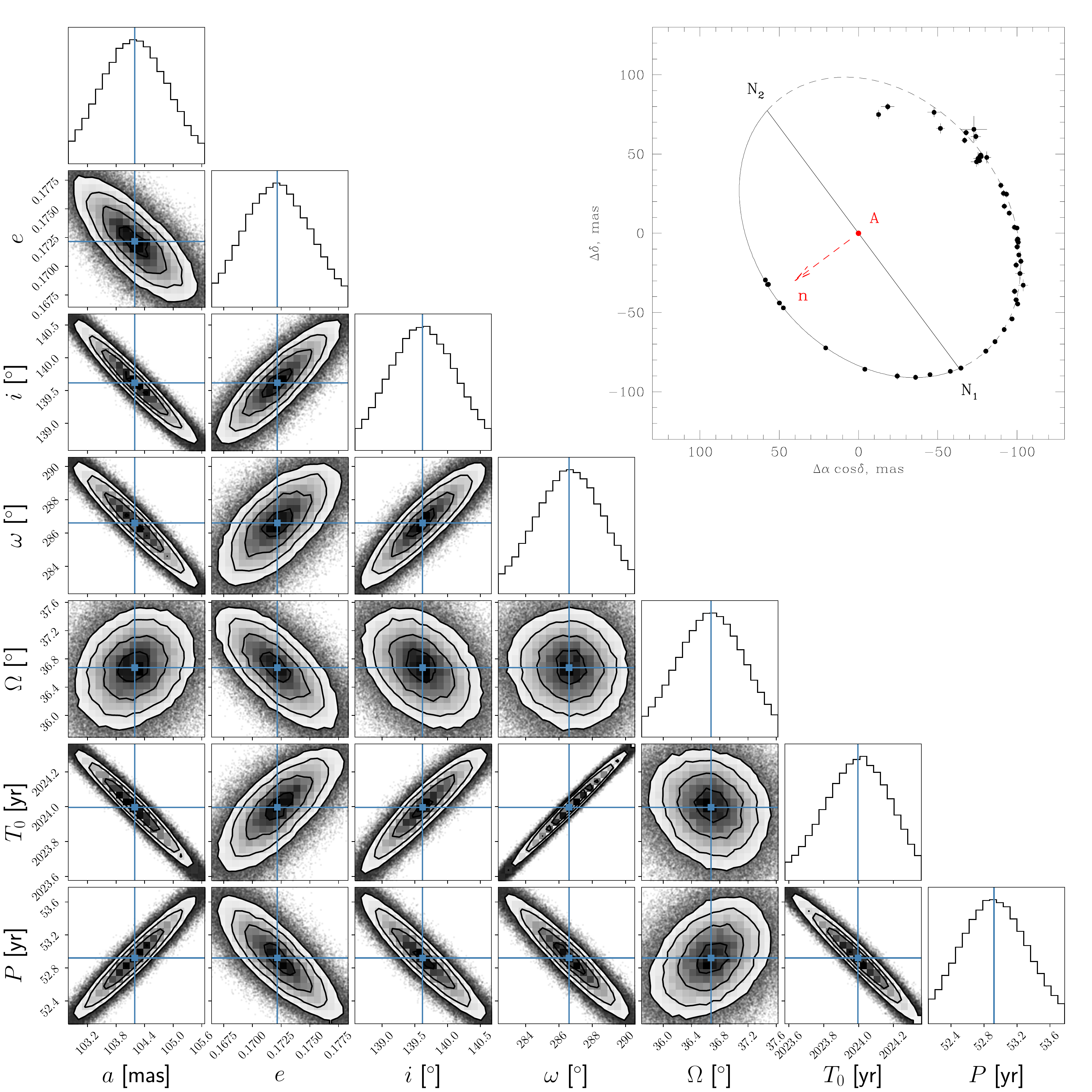}
      \caption{The lower-left panels show the marginal posterior probability distributions of the orbital parameters obtained with Method M2. The one-dimensional histograms along the diagonal (from top left to bottom right) represent the marginal distribution of each parameter. The off-diagonal panels, colored by intensity, display the two-dimensional marginal posterior density for each pair of parameters, with darker shades indicating higher probability density. The inset in the upper-right corner shows the orbit of the companion DF~Tau~B relative to the primary star DF~Tau~A. The solid and dashed lines indicate the portions of the orbit lying above and below the plane of the celestial sphere, respectively; thus, $N_1$ and $N_2$ denote the ascending and descending nodes of the orbit. The vector ${\bf n}$ indicates the orientation of the orbital angular momentum.
              }
         \label{fig:orbit}
   \end{figure*}

Our results are presented in Figure~\ref{fig:orbit} and Table~\ref{tab:orbit-params}: the orbital period $P$, time of periastron passage $T_0$, eccentricity $e$, semi-major axis $a$ (in angular units), inclination of the orbit to the line of sight $i$, position angle of the ascending node $\Omega$, argument of periastron $\omega$, and the total system mass $M_{\text{A}} + M_{\text{B}}$. For comparison, we also list the orbital parameters derived by \citet{Schaefer_2014}, \citet{Allen-2017}, and \citet{Kutra-2025}. In all cases, the total mass is scaled to a distance of $d = 140$~pc using the relation $M \propto (d/140)^3$.

%              ------------------
\begin{table*}
\renewcommand{\tabcolsep}{0.13cm}
\caption{Comparison of the orbital parameters of DF~Tau}
\label{tab:orbit-params}
 \begin{center}
  \begin{tabular}{c c c c c c c c c}
\hline
Ref.$^a$ & $P$ & $T_0$ & $e$ & $a$ & $180-i$ & $\Omega$ & $\omega$ & ${M_1+M_2}^b$ \\
& yr  &  yr   &     & mas & \textdegree & \textdegree  & \textdegree  & $M_\odot$ \\ 
\hline
S14 & $43.7 \pm 3.0$   & $1980.5 \pm 1.7$  & $0.287 \pm 0.067$ & $93.5 \pm 1.8$ 
& $35.7 \pm 2.1$ & $27.1 \pm 8.2$ & $302.6 \pm 7.6$  & $1.17 \pm 0.13$ \\ 
A17 & $46.1 \pm 1.9$   & $1979.2 \pm 1.5$  & $0.233 \pm 0.038$ & $94.9 \pm 2.2$ 
& $34.5 \pm 1.6$ & $33.9 \pm 5.0$ & $309.2 \pm 3.5$  & $1.10 \pm 0.12$ \\ 
K25 & $48.1 \pm 2.1$   & $1977.7 \pm 2.7$  & $0.196 \pm 0.024$ & $97.0 \pm 3.2$ 
& $54.3 \pm 2.4$ & $38.4 \pm 2.5$ & $310.6 \pm 9.2$  & $1.09 \pm 0.03$ \\ 
M1 & $52.9 \pm 0.8$ & $2024.6 \pm 0.1$ & $0.172 \pm 0.005$ & $104.2 \pm 1.3$ &
$40.4 \pm 1.0$ & $36.6 \pm 0.9$ & $286.7 \pm 3.8$ & $1.11 \pm 0.07$ \\
M2 & $52.9 \pm 0.4$ & $2024.0 \pm 0.2$ & $0.172 \pm 0.003$ & $104.2 \pm 0.7$ &
$40.4 \pm 0.5$ & $36.7 \pm 0.5$ & $286.6 \pm 2.1$ & $1.11 \pm 0.01$   \\
\hline
 \end{tabular}
  \end{center}
a) S14 -- \citet{Schaefer_2014}, A17 -- \citet{Allen-2017}, K25 -- \citet{Kutra-2025},  
M1 and M2 correspond to our results from Method 1 and Method 2, respectively;  b) $d=140$~pc;\\
\end{table*}
%              ------------------

Table~\ref{tab:orbit-params} shows that our results from the two methods are consistent with each other within their uncertainties, but differ significantly from those of \citet{Kutra-2025}. The most notable discrepancies concern the orbital period $P$ and the time of periastron passage $T_0$, which are essential for correlating the century-scale light curve with the changing separation between the DF~Tau components. The origin of this discrepancy remains unclear; however, for the purposes of this study, we adopt our own derived orbital parameters as the reference.

It is well known that the projection of an orbit onto the celestial sphere remains unchanged if the parameters $(\omega,\,\Omega)$ are simultaneously replaced by $(\omega - 180\degr,\,\Omega + 180\degr)$. Under this transformation, the radial velocity difference $\Delta V_{\text{r}} = V_{\text{r}}^{\text{B}} - V_{\text{r}}^{\text{A}}$ reverses sign, and the ascending and descending nodes of the orbit are interchanged \citep[see, e.g., Sect.~3.1]{Belinski-2022}. Taking this degeneracy into account and using the measured radial velocities of the components, $V_{\text{r}}^{\text{A}}$ and $V_{\text{r}}^{\text{B}}$, from \citet{Allen-2017}, we, like the authors of that study, conclude that the ascending node of the DF~Tau orbit lies in the southwest direction -- see Figure~\ref{fig:orbit}.

%                          ---------------------
%                          ---------------------

At a distance of $d = 140$~pc to DF~Tau, the semi-major axis of the system is $a \approx 14.6$~au, with the minimum and maximum separation between the components being 12.1 and 17.0~au, respectively. 

We note that \citet{Kalscheur-2025}, in determining the radial extent of the disk region where the disk wind of DF~Tau forms and produces emission in molecular hydrogen lines, derived a value of $R_{\text{H}_2} \approx 2.2$~au. As \citet{Kalscheur-2025} assumed a disk inclination of $i = 85\degr$ and that $R_{\text{H}_2} \propto \sin^2 i$, we scale their derived radius to our measured inclination of $i = 40\degr$, yielding $R_{\text{H}_2} \approx 1$~au.

                            %%%%%%%%%%%%%%%%%%%%%%%%%%%%%%

\subsection{The historical light curve of DF~Tau}  
\label{sec:lcurve}

\begin{figure*}
   \centering
   \includegraphics[width=\hsize]{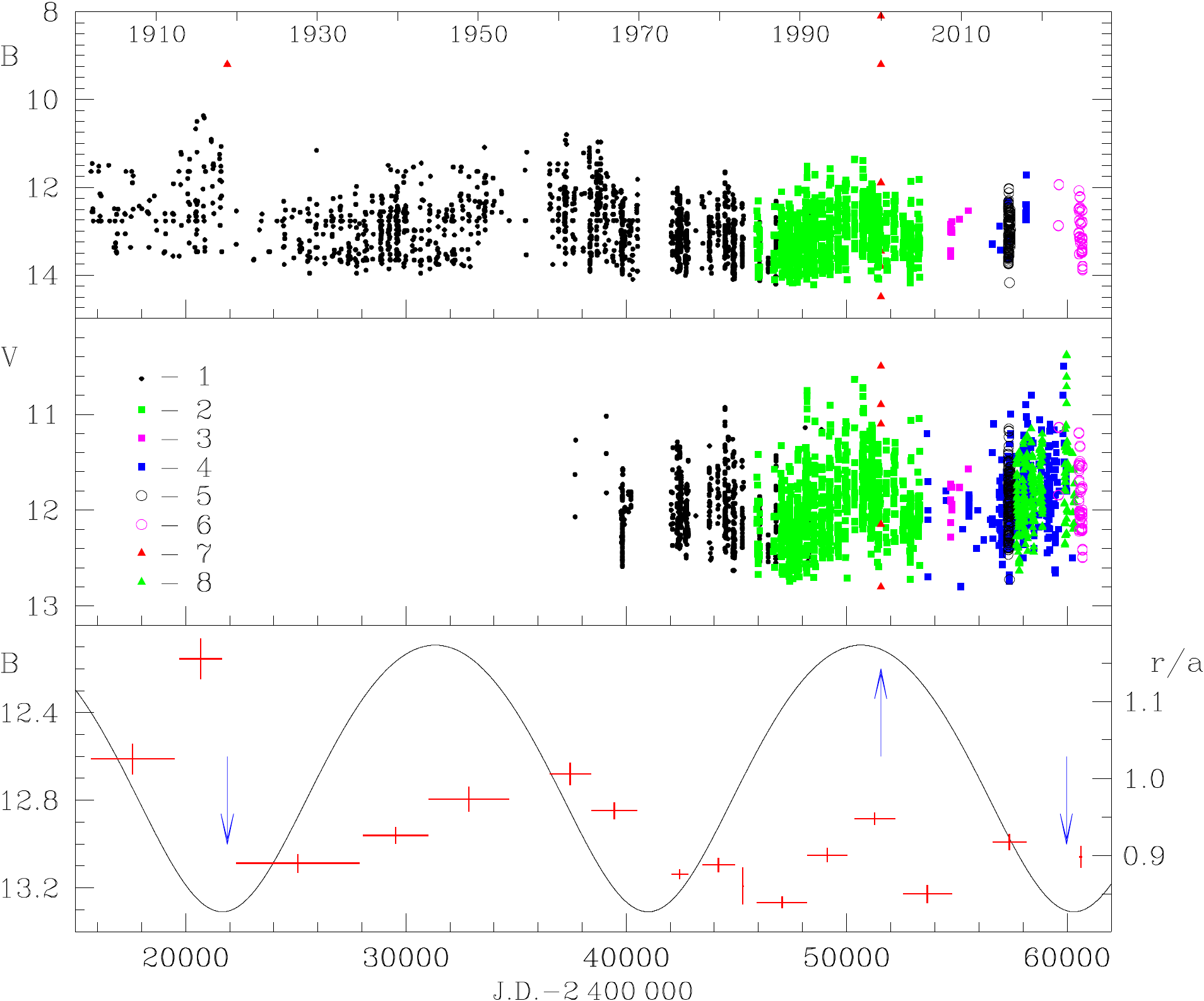}
\caption{Historical light curve of DF~Tau in the $B$ and $V$ bands (upper and middle panels, respectively). Data points are colored and symbol-coded according to the source: (1) \citet{Lamzin-2001}, (2) \citet{Grankin-07}, (3) our observations at the Crimean Astrophysical Observatory (CrAO), (4) AAVSO, (5) \citet{Allen-2017}, (6) our observations at CMO, (7) outbursts in 1918 \citep{Tsesevich-1973} and 2000 \citep{Li-2001}, and (8) \citet{Kutra-2025}. The lower panel shows the time evolution of the component separation, normalized to the system's semi-major axis. Red crosses mark the mean $B$-band brightness within each time interval, excluding the 1918 and 2000 outbursts. Blue arrows indicate the epochs of the 1918, 2000, and 2023 outbursts. See text for details.} 
    \label{fig:lcurve}
   \end{figure*}

In constructing the historical light curve of DF~Tau presented in Figure~\ref{fig:lcurve}, we combined our own measurements (described in Section~\ref{sect:observation}) with data from \citet{Lamzin-2001}, \citet{Tsesevich-1973}, \citet{Grankin-07}, \citet{Li-2001}, the \citet{Herbst-94} database, and the AAVSO archive. Additionally, T.~Kutra kindly provided us with $V$-band photometry of DF~Tau obtained with the 0.7~m and 1.1~m telescopes at Lowell Observatory, which were used in Figure~2 of \citet{Kutra-2025}.

%                          ---------------------
%                          ---------------------

As noted in the Introduction, \citet{Lamzin-2001} previously attempted to correlate the century-scale light curve of DF~Tau with the changing component separation, using the orbital parameters from \citet{Thiebaut-1995}. However, current measurements show that the orbital period and time of periastron passage differ significantly from those values. We therefore revisit the influence of component separation on the rate of accretion from the circumstellar (CS) disks, which governs the brightness of the hot (accretion) spot. Given that the effective temperatures of the DF~Tau components are $< 3700$~K, variations in the accretion rate are expected to most strongly affect the short-wavelength part of the spectrum.

Using our derived orbital parameters (Section~3.1), we show the predicted variation of the separation $r$ between the DF~Tau components as a function of time $t$ in the lower panel of Fig.~\ref{fig:lcurve}. Additionally, for 16 time intervals, the panel displays the mean $B$-band brightness $\bar{B}$ and the mean uncertainty of the mean $\bar{\sigma}_B$. The time intervals were chosen such that each contains sufficient measurements to yield a meaningful estimate of $\bar{\sigma}_B$, while also capturing characteristic features of the $B$-band light curve. This selection is somewhat arbitrary but allows a clearer view of the cyclic variation of $\bar{B}$ with time, revealing three local maxima of differing amplitude at rJD $\approx 20\,500$, $\approx 36\,000$, and $\approx 50\,500$. The intervals between these maxima are $\approx 41$~yr, shorter than the system's orbital period ($P \approx 53$~yr), indicating that the maxima occur at different orbital phases. As shown in the figure, the first maximum coincides with the companion nearing periastron, the second with the companion significantly before periastron, and the third nearly coincides with the time of apastron passage.
  
Strong $(\Delta B > 5^{\text{m}})$ brightenings of DF~Tau were observed on 24~October 1918 \citep{Tsesevich-1973} and 8--9~January 2000 \citep{Li-2001}, which we refer to as outbursts hereafter. As shown in Figure~\ref{fig:lcurve}, these outbursts occurred during the corresponding maxima in the light curve and are separated by $1.54 \pm 0.01$ orbital periods: the 1918 outburst occurred slightly before periastron passage, while the 2000 outburst occurred shortly after the maximum component separation.

A further outburst, detected only in the $V$ band, was recorded at the end of January 2023 (rJD $\approx 59970.6$) \citep{Kutra-2025}. Its amplitude exceeded $1^{\text{m}}$, at least twice the $\Delta V$ of the 2000 outburst; thus, we infer that the 2023 outburst was at least as luminous in the $B$ band as the two previous events. As shown in Figure~\ref{fig:lcurve}, this outburst occurred when the companion was nearing periastron, similar to the 1918 outburst.

%                          ---------------------
   \begin{figure}
   \centering
   \includegraphics[width=\hsize]{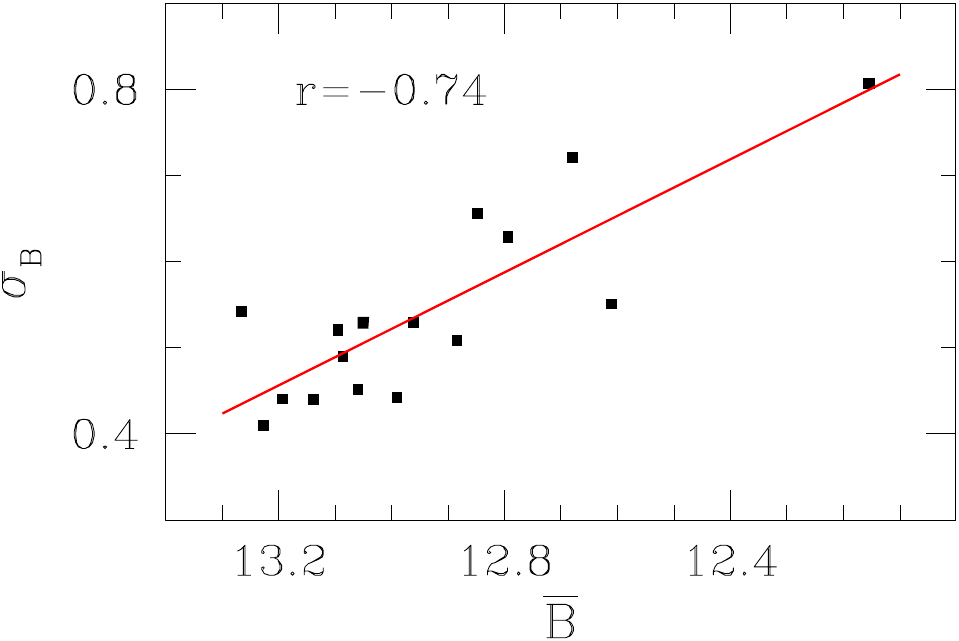}
\caption{The dependence of the standard deviation $\sigma_B$ on the mean brightness $\bar{B}$ for DF~Tau, derived from the time intervals indicated in the lower panel of Figure~\ref{fig:lcurve}, excluding the 1918 and 2000 outbursts. The red line shows a least-squares fit to the data; the correlation coefficient $r$ between $\sigma_B$ and $\bar{B}$ is also indicated. See text for details.
}
         \label{fig:correlation}
   \end{figure}
%                          ---------------------

Figure~\ref{fig:correlation} shows that as the mean brightness $\bar{B}$ decreases, so does the amplitude of variability $\sigma_B$. Throughout the entire observational baseline, the brightness of DF~Tau never dropped below $B \approx 14\fm1$. \footnote{The only exception is a single measurement by \citet{Li-2001}, taken a few days after the 2000 outburst, when the brightness fell to $B = 14\fm5 \pm 0\fm2$; this is discussed in the following section.}  
This is likely due to the fact that the century-scale and seasonal photometric variability of DF~Tau is dominated by the primary \sout{star }DF~Tau~A. This interpretation is consistent with differential photometry of the system, which shows that on 27~July 1994, the companion was $1\fm3$ fainter than the primary in the $B$ band, with a measured brightness of $14.47 \pm 0.13$ \citep{White-Ghez-2001}.

\begin{figure}
   \centering
   \includegraphics[width=0.7\hsize]{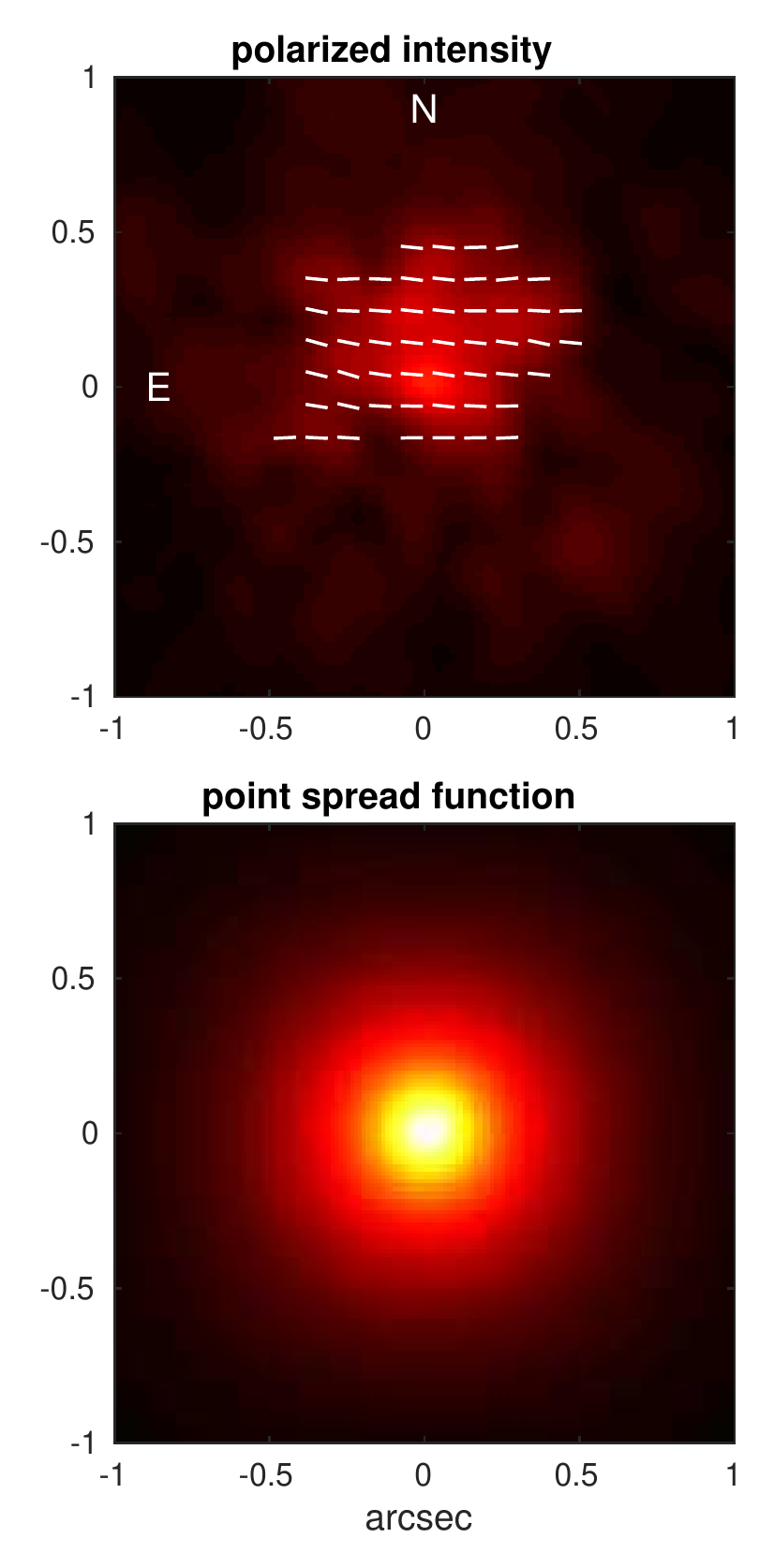}
\caption{\textit{Top:} Image of DF~Tau in polarized intensity, reconstructed using differential speckle polarimetry. \textit{Bottom:} Point spread function used in the reconstruction (shown for illustration of angular resolution).
}
         \label{fig:DSP}
   \end{figure}

Our polarimetric observations in the $I$ band (Section~\ref{sect:observation}) yield a degree of polarization $p_{\text{I}} = 0.37 \pm 0.15$ and a polarization angle $\theta_{\text{I}} = 94^{\circ} \pm 24^{\circ}$, which are consistent within uncertainties with the mean values reported by \citet{Shakhovskoj-2006} from observations between October 1991 and November 1998. The distribution of polarized intensity, reconstructed using differential speckle polarimetry \citep{Safonov-2019}, is shown in Figure~\ref{fig:DSP}.

%                          ---------------------
%                          ---------------------

\begin{figure}
   \centering
   \includegraphics[width=\hsize]{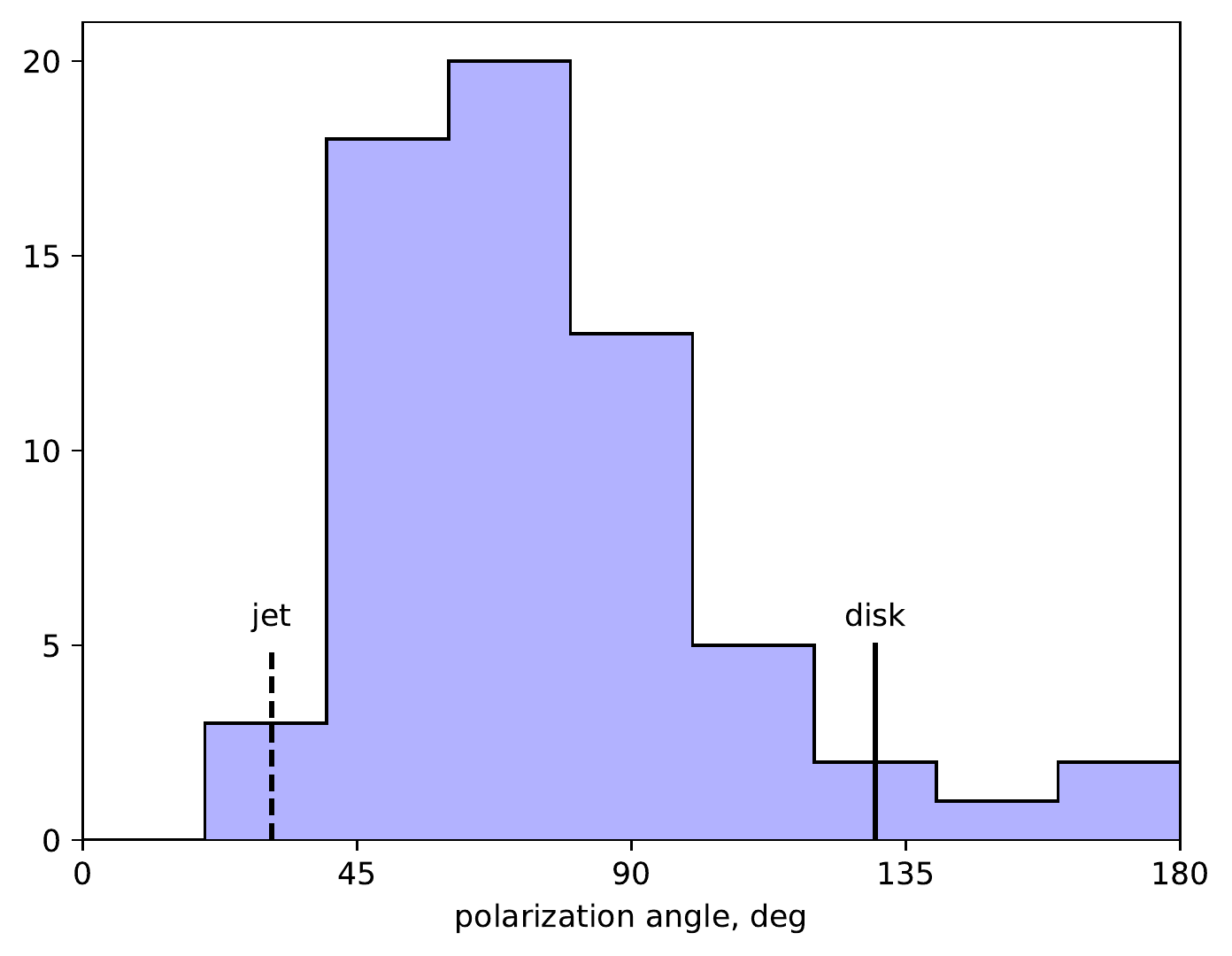}
   \caption{Histogram of the polarization angle distribution of DF~Tau, constructed from the data of \citet{Shakhovskoj-2006}. The dashed line shows the expected polarization orientation from the jet. The solid line shows the expected polarization orientation from the disk. 
   }
         \label{fig:polarimetry}
   \end{figure}

Figure~\ref{fig:polarimetry} presents a histogram of the distribution of polarization angles $\theta_{\text{I}}$ based on the data of \citet{Shakhovskoj-2006}. More than 75\% of all measurements fall within the range $40\degr$ to $100\degr$ -- as does our measurement. A similar result is obtained for polarization angles in the $UBVR$ bands, indicating that the mean polarization angle $\theta$ is nearly independent of wavelength across the optical range. The same angular range ($40\degr$–$100\degr$) is also observed in polarization measurements obtained with narrowband filters in October 1976 \citep{Bastien-1982} and in October 1984 and December 1985 \citep{Bastien-Menard-1992}. We therefore conclude that orbital motion does not produce a systematic variation in the polarization angle.

%                          ---------------------
%                          ---------------------

                            %%%%%%%%%%%%%%%%%%%%%%%%%%%%%%

\section{Discussion}
\label{sec:discuss}

The irregular photometric variability of CTTS on a wide range of timescales may arise not only from changes in the rate of accretion onto the central star, but also from variable circumstellar extinction, as in the case of RW~Aur~A (see \citealt{Dodin-2019} and references therein). According to our orbital solution (Table~\ref{tab:orbit-params}), the inclination of the DF~Tau disk axis to the line of sight is $i \approx 40\degr$. Therefore, if the variability is due to extinction, it is most likely caused by occultations of DF~Tau~A by inhomogeneities in the dusty disk wind. \footnote{Unlike BP~Tau ($i=38\degr$), no sharp dimming events have been observed in DF~Tau, making the scenario of dust clumps in the magnetosphere obscuring the star \citep{Burlak-2025} less plausible.}  
However, in such a case, the magnitude of the effect is unlikely to be significant, given the low extinction toward DF~Tau: $A_\text{V} < 0.5^{\text{m}}$ \citep{Herczeg-14} and $N_\text{H} < 10^{21}$~cm$^{-2}$ \citep{McJunkin-Lya-2014}. In other words, we agree with the conclusions of \citet{Shakhovskoj-2006} and \citet{Kutra-2025} that the photometric variability of DF~Tau is predominantly driven by changes in the rate of accretion onto the primary component.

When a binary system is formed from a rotating protostellar cloud, a triple-disk configuration generally arises: two CS disks around each star and an outer CB disk -- see the review by \citet{Offner-PPVII} and references therein. Tidal interactions cause the outer radii of the CS disks to shrink to $\sim 0.4a$ and the inner radius of the CB disk to expand to $\sim 2a$, where $a$ is the semi-major axis of the binary system, within timescales $\lesssim 10^3$ orbital periods $P_{\text{orb}}$ \citep{Papaloizou-Pringle-1977, Artymowicz-Lubow-1994}. In this configuration, one or two spiral gas streams extend from the inner edge of the CB disk, feeding material to the CS disks of the individual components -- see, for example, Figure~1 in \citet{Rosotti-Clarke-2018} and Figures~3-4 in \citet{Bisikalo-2012}.

A similar configuration is observed in the young binary system DQ~Tau, with an orbital period of $15\fd8$ and an eccentricity of $0.57$ \citep{Tofflemire-2017}. This system exhibits quasiperiodic photometric variations driven by modulations in the accretion rate $\dot{M}_{\text{acc}}$, with a period equal to the orbital period and maxima occurring near periastron \citep{Jensen-2007, Tofflemire-2017, Tofflemire-2025}. This phenomenon, termed ``pulsating accretion'' by \citet{Jensen-2007}, is attributed to enhanced mass transfer from the CB disk to the CS disks of the components during closest approach, in agreement with numerical simulations by \citet{Muñoz-Lai-2016}.

However, short outbursts (increases in $\dot{M}_{\text{acc}}$) have been repeatedly observed in DQ~Tau when the companion is near apastron -- see \citet{Tofflemire-2017} and references therein. Given that the companion lies closest to the CB disk at apastron, \citet{Bary-2014} proposed that these apastron outbursts result from interactions between the companion and a tidal arm of the CB disk. However, numerical simulations by \citet{Muñoz-Lai-2016} do not support this scenario.

On the other hand, a significant number of young binary systems exhibit only one or two CS disks without a CB disk -- see \citet{Cuello-2025} and references therein. As noted in the Introduction, DF~Tau belongs to this class: both components are surrounded by CS disks, but no CB disk has been detected \citep{Kutra-2025}. Numerical simulations of such systems show that, in the presence of a significant orbital eccentricity, the accretion rate undergoes periodic enhancement near the time of periastron passage by the companion -- similar to the behavior observed in systems with a CB disk (see \citealt{Picogna-Marzari-2013} and references therein).

This theoretical prediction is supported by the high brightness level of DF~Tau and the 1918 outburst, which occurred near the time of periastron passage by the companion. The 2023 outburst also coincided with the companion being near periastron, yet the seasonal mean brightness was relatively low. However, as noted in Section~3.2, the maximum brightness following the 1918 event occurred when the companion was approximately midway between periastron and apastron, and the 2000 outburst and its associated brightness peak coincided with the moment of maximum component separation. This pattern of accretion rate modulation with orbital phase is inconsistent with the theoretical prediction.

Thus, current calculations can explain neither the absence of periodic accretion rate modulation in systems without a CB disk (DF~Tau) nor the brief enhancement of $\dot{M}_{\text{acc}}$ near apastron in systems with a CB disk (DQ~Tau).
 
The nearest observations following the 1918 outburst are separated by more than 200~days \citep{Tsesevich-1973}, whereas the 2000 and 2023 outbursts lasted no more than two days \citep{Li-2001}. Therefore, it is quite possible that similar short-lived outbursts occurred between 1905 and 2024 but were missed. In principle, the jet of DF~Tau could provide evidence for such events.

We recall that jets from young stars are bipolar outflows of gas extending up to several parsecs, moving at velocities of several hundred \kms{} relative to their surroundings \citep{Bally-2016}. These jets appear as chains of compact emission nebulae -- known as Herbig--Haro (HH) objects -- embedded in a more diffuse and faintly emitting gas. HH objects are associated with shock fronts generated either by the collision of a hypersonic outflow with the ambient medium or by faster material within the jet overtaking slower gas ejected previously \citep{Raga-1990}. The emergence of such high-velocity flows is naturally linked to transient increases in the accretion rate, which should be accompanied by photometric outbursts. Consequently, estimating the age of an individual HH object provides information on the epoch of the corresponding photometric outburst.

According to \citet{Li-2001}, following the sharp increase in the accretion rate during the 2000 outburst of DF~Tau, a mass ejection with a velocity of 200--300~\kms{} occurred. The ejected shell rapidly shielded DF~Tau~A, causing the combined brightness of the system to drop below the minimum level observed before or after the event, and transforming emission lines at wavelengths $\lambda < 0.44~$\micron{} into absorption ones with P~Cygni-type profiles. However, no new HH object appeared in the jet ($\text{PA}_{\text{j}} \approx 301\degr$) or counter-jet ($\text{PA}_{\text{cj}} \approx 133\degr$) at the expected distance from DF~Tau -- see the right panel of Figure~1 in \citet{Dodin-2025}.

However, the same figure reveals a weakly collimated outflow at a position angle of $\approx 270\degr$ and a distance of $\approx 5.4^{\prime\prime}$ from DF~Tau, whose dynamical timescale is consistent with the 2000 outburst. No other HH objects with dynamical timescales corresponding to outbursts between 1918 and 2000 are detected in either the jet or the counter-jet.

We also note that numerical simulations of jet formation in young binary systems \citep{Sheikhnezami-2024, Sheikhnezami-2025} predict that a spiral density wave arising in the CS disk should be visible in the jet. However, no such structures have been detected in the jet or counter-jet of DF~Tau \citep{Dodin-2025}, although deeper H$\alpha$ imaging may reveal them.

The dominant polarization direction of DF~Tau is oriented along PA $\approx 70\degr$ (see Figure~\ref{fig:polarimetry}), which is inconsistent with the expected scattering direction from either the CS disk of DF~Tau~A or the dusty wind blowing perpendicular to the disk \citep{Dodin-2019, Shulman-Grinin-2019}. Evidently, the polarization-producing dust structure lies along PA $\approx 160\degr$ or $340\degr$, and the scattering dust has a characteristic size $r_{\text{d}} \lesssim 0.1~$\micron. Indeed, the degree of polarization of DF~Tau increases monotonically with decreasing wavelength, reaching $\sim 2{-}3\,\%$ in the $U$ band \citep{Shakhovskoj-2006}. The wavelength $\lambda_{\text{max}}$ corresponding to the maximum polarization is related to the characteristic dust grain radius by $r_{\text{d}} \approx \lambda_{\text{max}}/4$ \citep{vanDeHulst1957}.

Figure~1 of \citet{Dodin-2025} shows that a diffuse reflection nebula lies $\approx 60^{\prime\prime}$ to the southeast of DF~Tau, marking the edge of the dark cloud B213 \citep{Onishi-1996}. In the immediate vicinity of the star, at angular distances of $5^{\prime\prime}$--$10^{\prime\prime}$, small-scale reflection nebulae are also visible. However, our polarimetric observations (Figure~\ref{fig:DSP}) indicate that the source of polarized light lies within $\lesssim 0.5^{\prime\prime}$ of DF~Tau. The polarization parameters show no variation with orbital phase, making it unlikely that the scattering material is located at distances comparable to the semi-major axis of the orbit ($\sim 0.1^{\prime\prime}$). Instead, the scattering responsible for the observed polarization must occur on dust structures much closer to the star than $0.1^{\prime\prime}$. This conclusion is further supported by the short timescale of polarization variability reported by \citet{Shakhovskoj-2006}.

                        %%%%%%%%%%%%%%%%%%%%%%%%%%%%%%

\section{Conclusion}
\label{sec:conclusion}

We have redetermined the orbital parameters of the young binary system DF~Tau using two independent methods that yield nearly identical results. The most significant updates are the orbital period $P \approx 52.9$~yr and the time of periastron passage $T_0 \approx 2024.3$. These refined parameters enable us to investigate the dependence of the system's photometric activity on the component separation, which varies between 12.1 and 17.0~au according to our analysis.

The discrepancy between the theoretically predicted and observed modulation of the mean brightness (accretion rate) due to orbital motion in DF~Tau is intriguing in its own right. However, the fact that a similar discrepancy between theory and observation is also observed in other young binary systems \citep{Jensen-2007} suggests that the influence of binary companions on CS disks remains poorly understood. This is not surprising given the complexity of the problem (see, e.g., the Introduction in \citealt{Kurbatov-2017}), but it clearly underscores the need for new observational and theoretical investigations in this area.

DF~Tau itself also warrants further investigation. It remains unclear how to explain the presence of an intense, quasi-spherical wind alongside collimated outflows in which the jet and counter-jet are not aligned along opposite directions \citep{Dodin-2025}. It is also unclear why the 2000 outburst produced a weakly collimated outflow whose direction differs from that of the jets. High-resolution imaging of DF~Tau is strongly desirable to resolve the structure of the CS disks and to refine their inclination relative to the orbital plane -- in particular, to explain the non-trivial morphology of the HH objects in the vicinity of the system.

                        %%%%%%%%%%%%%%%%%%%%%%%%%%%%%%
                        
\section*{Acknowledgements}

We thank the referee for helpful comments, the staff of CMO SAI MSU for assistance with observations, T.~V. Demidova and S.~Yu. Parfenov for valuable discussions, Prof.~W.~Herbst for providing access to his photometric database \citep{Herbst-94}, and T.~Kutra for sharing photometric data obtained at Lowell Observatory (USA). We also acknowledge the use of the SIMBAD database (CDS, Strasbourg, France), the Astrophysics Data System (ADS, NASA, USA), and the AAVSO database (https://www.aavso.org).

                        %%%%%%%%%%%%%%%%%%%%%%%%%%%%%%
    
\section*{Financial support}

The study was conducted under the state assignment of Lomonosov Moscow State University. The results were obtained using equipment acquired under the Lomonosov Moscow State University Program of Development.

                        %%%%%%%%%%%%%%%%%%%%%%%%%%%%%%
\section*{Conflict of interest}
The authors declare no conflict of interest.

                        %%%%%%%%%%%%%%%%%%%%%%%%%%%%%%
\bibliographystyle{aspb1}
\bibliography{dftau}

                        %%%%%%%%%%%%%%%%%%%%%%%%%%%%%%

\section*{Appendix}
\label{sec:appendix}

%              ------------------
\begin{table*}
\renewcommand{\tabcolsep}{0.18cm}
\caption{Astrometry of DF~Tau}
\label{tab:orbit-dat}
 \begin{center}
  \begin{tabular}{lccccc|lccccc}
\hline
Date      & $\rho$ & $\sigma_{\rm \rho}$ & PA      & $\sigma_{\rm PA}$ & Ref. &
Date      & $\rho$ & $\sigma_{\rm \rho}$ & PA      & $\sigma_{\rm PA}$ & Ref. \\
yr        & mas    &  mas                & $\degr$ & $\degr$           &      &    
yr        & mas    &  mas                & $\degr$ & $\degr$           & \\
\hline
1986.80   &    76   &    3    &   350.4  &  0.5  &   [1] &  2001.164   &   100.5  &    2.0  &   265.1   &   1.2    &  [5] \\
1989.840  &   82    &    2    &   347.0  &  3.0  &   [1] &  2001.92    &   104    &    2    &   260.2   &   1      &  [8] \\
1990.857  &   90.0  &    2    &   328.0  &  3.0  &   [2] &  2002.129   &   102.1  &    2.0  &   262.3   &   1.2    &  [5] \\
1991.730  &   84    &    4    &   322.0  &  1.0  &   [1] &  2002.74    &   105    &    3    &   256.0   &   2      &  [8] \\
1992.778  &   98.0  &    8    &   312.0  &  5.0  &   [2] &  2003.061   &   101.5  &    2.0  &   258.6   &   1.2    &  [8] \\
1993.734  &   89.0  &    2.0  &   311.2  &  1.3  &   [3] &  2003.85    &   109    &    3    &   252.5   &   2      &  [8] \\
1993.816  &   93.0  &    2.0  &   313.1  &  1.3  &   [3] &  2003.883   &   105.1  &    2.0  &   249.6   &   1.1    &  [8] \\
1993.901  &   96.0  &    4    &   309.5  &  0.7  &   [2] &  2004.8215  &   108    &    2    &   247.1   &   0.5    &  [9] \\
1994.569  &   87.1  &    3.8  &   301.2  &  2.0  &   [4] &  2004.9815  &   109.93 &   0.49  &   246.06  &   0.26   & [10] \\
1994.797  &   89.0  &    2    &   302.0  &  3.0  &   [2] &  2005.9316  &   110.92 &   1.59  &   240.84  &   0.82   & [10] \\
1994.821  &   91.2  &    2.0  &   302.1  &  1.3  &   [3] &  2006.9631  &   110.22 &   0.59  &   236.52  &   0.31   & [10] \\
1994.940  &   94    &    2    &   300.6  &  2.5  &   [1] &  2008.0445  &   109.97 &   0.35  &   231.57  &   0.18   & [10] \\
1994.966  &   89.0  &    1    &   301.0  &  1.0  &   [2] &  2008.9618  &   109.54 &   0.29  &   227.21  &   0.15   & [10] \\
1995.055  &   91.6  &    2.0  &   302.6  &  1.3  &   [3] &  2011.0645  &   106.88 &   0.35  &   217.24  &   0.19   & [10] \\
1995.572  &   90.7  &    2.0  &   302.3  &  1.3  &   [3] &  2011.7798  &   104.63 &   0.98  &   213.63  &   0.54   & [10] \\
1997.019  &   94.8  &    2    &   288.6  &  1.3  &   [5] &  2013.0740  &    99.97 &   1.35  &   206.84  &   0.78   & [10] \\
1997.706  &   94.9  &    2    &   285.4  &  1.3  &   [5] &  2014.0679  &    97.80 &   0.66  &   201.61  &   0.39   & [10] \\
1997.884  &   96.6  &    2    &   284.8  &  1.3  &   [5] &  2015.0017  &    93.39 &   2.24  &   195.20  &   1.37   & [11] \\
1998.164  &   93.6  &    2    &   280.5  &  1.3  &   [5] &  2016.8028  &    85.89 &   0.41  &   182.67  &   0.27   & [11] \\
1998.775  &   96    &    2    &   277.6  &  1.2  &   [6] &  2019.0521  &    75.27 &   0.59  &   164.05  &   0.45   & [12] \\
1999.695  &   98.5  &    2.0  &   272.2  &  1.2  &   [5] &  2021.647   &    66.82 &   0.12  &   134.82  &   0.081  & [13] \\
1999.819  &  100.0  &    1.0  &   271.9  &  0.6  &   [7] &  2021.828   &    66.53 &   0.45  &   131.40  &   0.32   & [13] \\
2000.241  &  100.4  &    2.0  &   267.5  &  1.2  &   [5] &  2022.7981  &    65.39 &   1.14  &   119.53  &   1.0    & [12] \\
2000.671  &  100.9  &    2.0  &   266.8  &  1.2  &   [5] &  2022.849   &    65.88 &   0.23  &   119.41  &   0.26   & [13] \\
2001.063  &  100.5  &    2.0  &   267.9  &  1.2  &   [5] &  2022.9919  &    65.73 &   0.9   &   116.68  &   0.78   & [12] \\
\hline
 \end{tabular}
  \end{center}
1 -- \citet{Thiebaut-1995}; 2 -- \citet{Ghez-1995}; 3 -- \citet{Simon-1996};
4 -- \citet{Ghez-1997}; 5 -- \citet{Schaefer-2003}; 6 -- \citet{Balega-2002};
7 -- \citet{Balega-2004}; 8 -- \citet{Schaefer-2006}; 9 -- \citet{Balega-2007};  
10 -- \citet{Schaefer_2014}; 11 -- \citet{Allen-2017}; 12 -- \citet{Kutra-2025}; 
13 -- \citet{Grant-2024}.
\\
\end{table*}
%              ------------------

        %%%%%%%%%%%%%%%%%%%%%%%%%%%%%%%%%%%%%%%%%%%%%
    
%\bibliographystyle{aspb1}
%\bibliography{dftau}

\end{document}

%% file: sao_cmd_author.tex
\def\squareforqed{\hbox{\rlap{$\sqcap$}$\sqcup$}}

\def\sq{\ifmmode\squareforqed\else{\unskip\nobreak\hfil
\penalty50\hskip1em\null\nobreak\hfil\squareforqed
\parfillskip=0pt\finalhyphendemerits=0\endgraf}\fi}

\def\degr{\hbox{$^\circ$}}

\def\utw{\smash{\rlap{\lower5pt\hbox{$\sim$}}}}

\def\udtw{\smash{\rlap{\lower6pt\hbox{$\approx$}}}}

\def\fd{\hbox{$\,.\!\!^{\rm d}$}}

\def\fm{\hbox{$\,.\!\!^{\rm m}$}}

\def\diameter{{\ifmmode\mathchoice
{\ooalign{\hfil\hbox{$\displaystyle/$}\hfil\crcr
{\hbox{$\displaystyle\mathchar"20D$}}}}
{\ooalign{\hfil\hbox{$\textstyle/$}\hfil\crcr
{\hbox{$\textstyle\mathchar"20D$}}}}
{\ooalign{\hfil\hbox{$\scriptstyle/$}\hfil\crcr
{\hbox{$\scriptstyle\mathchar"20D$}}}}
{\ooalign{\hfil\hbox{$\scriptscriptstyle/$}\hfil\crcr
{\hbox{$\scriptscriptstyle\mathchar"20D$}}}}
\else{\ooalign{\hfil/\hfil\crcr\mathhexbox20D}}%
\fi}}